\documentclass[twocolumn,showpacs,preprintnumbers,amsmath,amssymb]{revtex4}
\usepackage{graphicx}
\usepackage{dcolumn}
\usepackage{bm}
\usepackage{color}
\usepackage{subfigure}

\begin{document}

%%\preprint{APS/123-QED}
\title{Ultracold Thermalization of $^7$Li and $^{87}$Rb}

\author{C. Marzok}
\author{B. Deh}
\author{Ph.W. Courteille}
\author{C. Zimmermann}
\affiliation{Physikalisches Institut, Eberhard-Karls-Universit\"at
T\"ubingen,
\\Auf der Morgenstelle 14, D-72076 T\"ubingen, Germany}

\date{\today}

\begin{abstract}
We report on measurements of cross-species thermalization inside a
magnetically trapped spin-polarized mixture of $^{87}$Rb and
$^7$Li atoms with both atoms in their respective low field seeking
magnetic substates $\left|F=2,m_F=2\right>$. Measurement of the
thermalization velocity in the ultracold regime below $10\,\mu$K
allows for the derivation of the absolute value of the pure triplet
s-wave scattering length governing the interaction. We find
$|a_{7,87}|=(59\pm19)\,a_{\rm B}$. We propose to study both species in
the condensed regime to derive the sign of $a_{7,87}$. In this
context, we present numerical solutions to the coupled
Gross-Pitaevskii equation based on the hypothesis of a positive sign. According to the
simulations, phase separation of the Li and Rb
$\left|2,2\right>$ clouds occurs along with a mean-field stabilization
allowing for larger atom numbers of condensed $^7$Li atoms before
collapse sets in. Observation of this mean-field stabilization
would directly fix the sign of $a_{7,87}$. We discuss our results in
the light of this proposal.
\end{abstract}


\maketitle
\section{Introduction}
Sympathetic cooling with an independent heat bath is a convenient
tool to cool atomic clouds that cannot be cooled directly. Its
application to Fermi-Bose mixtures led to a breakthrough in
reaching the quantum degenerate regime with fully spin polatized fermionic atom clouds.
The antisymmetry of the fermionic wave functions prohibits direct
evaporative cooling due to missing s-wave scattering. Adding
a second bosonic species that can have collisions with the
fermionic species circumvents this difficulty. A range of
different Fermi-Bose mixtures was brought to the degenerate regime
employing this technique: $^{6}$Li+$^{7}$Li \cite{Truscott:01},
$^{6}$Li+$^{23}$Na \cite{Hadzibabic:02}, $^{40}$K+$^{87}$Rb
\cite{Roati:02},
 $^{6}$Li+$^{87}$Rb \cite{Silber:05}, $^3$He+$^4$He \cite{McNamara:06}. For bosonic target
 species it is also interesting to be cooled by an independent cooling agent. Two $^{87}$Rb condensates in different hyperfine states $F$ could be
 produced by evaporation of atoms in the $F=2$ state while atoms in $F=1$ were passively cooled through thermalizing collisions \cite{Myatt:96}.
Another example is $^{85}$Rb, whose intra-species scattering
length $a_{85}$ has a zero crossing at collision energies
$E_{coll}/k_B\sim375~\mu$K. This prohibits direct evaporation of
$^{85}$Rb starting with higher temperatures.
 The problem was overcome by sympathetically cooling $^{85}$Rb in a bath of $^{87}$Rb \cite{Bloch:01}.
 Sympathetic cooling down to quantum degeneracy of the target
 species was also realized in $^{41}$K+$^{87}$Rb \cite{Modugno:01,Ferrari:02} and
recently in $^{39}$K+$^{87}$Rb \cite{Modugno:07} where again
$^{87}$Rb served as coolant. Furthermore, interspecies scattering
properties were determined in a mixture of $^{87}$Rb and
$^{133}$Cs, which was brought to low temperatures with the same
technique \cite{Anderlini:05}.
Mixtures of different isotopes of Yb \cite{Takasu:05} as well as a
$^{174}$Yb+$^{87}$Rb mixture \cite{Kroboth:07} are currently under
examination as well.

When working with $^7$Li, it is important to note that
Bose-Einstein condensates of this species are intrinsically
unstable in the $\left|F,m_F\right>=\left|2,2\right>$ hyperfine
state because the $^7$Li triplet scattering length
$a_7=-23.7\,a_{\rm B}$ is negative. This instability and the maximum
condensate size were predicted theoretically \cite{Ruprecht:95}
and later confirmed experimentally \cite{Sackett:97}. The strong
effects that negative scattering lengths have on condensates were
also shown in the controlled destabilization \cite{Roberts:01} and
the spectacular 'bose\-nova' experiments \cite{Donley:01} in
Bose-Einstein condensates of $^{85}$Rb, where the strength and
sign of the scattering length was rapidly changed by means of a Feshbach
resonance. Our interest in cooling $^7$Li with $^{87}$Rb
originates in the mean-field interaction in the regime
where both species are Bose-condensed.
Depending on the strength of this interaction, it can strongly
alter the stability conditions for $^7$Li Bose-Einstein
condensates, as we show in a theoretical section of this paper.

Here, we report on the experimental realization of sympathetic
cooling of $^7$Li in a bath of $^{87}$Rb. From controlled
thermalization experiments in a temperature range from 10$\,\mu$K
to 2$\,\mu$K we extract the interspecies scattering cross section
$\sigma_{7,87}=4\pi a_{7,87}^2$. This fixes the absolute value of
the triplet scattering length to $\left|a_{7,87}\right|=(59\pm19)\,a_{\rm B}$
. Upon cooling further down, we reach criticality with
respect to Bose-Einstein condensation in $^7$Li before $^{87}$Rb
starts to condense. The $^7$Li criticality manifests itself in
additional sudden losses of the $^7$Li atom number as for our trap conditions
the atom number of the overcritical $^7$Li largely exceeds the atom
number of stable $^7$Li condensates.

We also present numerical solutions of the coupled Gross-Pitaevskii equations of two
interacting condensates of $^7$Li and $^{87}$Rb in spherically
symmetric traps. Stabilization of the Li condensate for large Rb atom numbers is observed
for a positive sign of $a_{7,87}$. Experimentally, observation of this effect can be expected when
intra- and interspecies inelastic loss coefficients are small and the overlap is good.
A precondition for this is that the relative gravitational sag is smaller than the Thomas-Fermi-radii
of the condensates involved.
Hence, by observing the collapse of Li and the impact of
the mean-field energy of Rb on the collapse in a regime where both
species are condensed, it is, in principle, possible to determine
the sign of the mutual interaction \cite{Adhikari:01b}.

We structure the paper as follows: In Section \ref{sec:exp} we
shortly describe the apparatus and the experimental procedure used
for cooling. In Section \ref{sec:cross}, experimental results for
crossed species thermalization are presented and the absolute
value of the scattering length determined. Section \ref{sec:bec}
presents experimental evidence that we are able to reach the
critical temperature for Bose-Einstein condensation of $^7$Li.
Finally, Section \ref{sec:mean} describes the numerical
calculations for coupled $^7$Li and $^{87}$Rb condensates.

\section{Experimental setup}\label{sec:exp}

Our experimental setup is identical to the one described in
\cite{Silber:05} where we studied thermalization between fermionic $^6$Li and $^{87}$Rb,
except for the fact that we now use the bosonic isotope of Li.
As atom sources we use commercial dispensers with
(enriched) $^{87}$Rb and a Zeeman slower for the
$^7$Li. Up to $3\cdot10^8$ Rb and $8\cdot10^7$ Li atoms are
simultaneously trapped in two spatially overlapping
magneto-optical traps (MOT). We then compress the MOTs by
increasing the trap coil currents and then apply a short (1~ms)
stage of polarization gradient cooling. The Rb then has a
temperature of $T_{87}\approx80\,\mu$K. The Li, for which
polarization gradient cooling does not work because of the
unresolved hyperfine structure in the excited state, has an
estimated temperature of several hundred $\mu$K. Before turning on
the magnetic trap, the Rb is optically pumped into the stretched
state $\left|2,2\right>$ while Li is left
unpolarized. We magnetically trap almost
all Rb atoms and approximately $10^7$ Li atoms
\cite{interacting_mots}. The mixture is then adiabatically
transferred via several intermediate magnetic traps into a
Ioffe-Pritchard type trap characterized by the secular frequencies
$\tilde{\omega}_{87}/2\pi=(50\times206\times200)^{1/3}~$Hz for the
Rb $\left|2,2\right>$ state (for Li $\left|2,2\right>$ the frequencies are
higher by a factor $\sqrt{87/7}$) and the magnetic field offset
$3~$G. As in the case of $^6$Li, purity of the Rb spin states has
to be ensured by special measures. Directly after turning on the
Ioffe-Pritchard trap, a cleaning microwave sweep coupling residual
Rb $\left|1,-1\right>$ atoms to the untrapped $\left|2,-2\right>$
state is applied. The Rb cloud is then selectively cooled by
forced microwave evaporation coupling the $\left|2,2\right>$ state
with the anti-trapped $\left|1,1\right>$ state at a certain
distance from the trap centre. An additional microwave knive has
to be applied that constantly empties the trap from
$\left|2,1\right>$ Rb atoms by coupling them to the untrapped
$\left|1,0\right>$ state. The $\left|2,1\right>$ atoms accumulate
during the last stages of evaporation via dipolar relaxation
\cite{Guenther:06} and lead to severe losses in Li atom
numbers through inelastic collisions. The Li cloud, which is not
actively cooled, adjusts its temperature to the Rb cloud via
interspecies thermalization. We find that $^7$Li thermalizes much
faster than its fermionic counterpart $^6$Li \cite{Silber:05}.
Atom numbers and temperatures are derived by
standard optical absorption imaging. The trap is rapidly turned
off and the clouds expand ballistically for a variable amount of
time. Light pulses resonant with the $F=2\rightarrow F'=3$
transitions of the respective $D_2$-lines then map the density
distribution onto CCD cameras. In the case of Li, an additional
counter-propagating repumping beam resonant with the
$F=1\rightarrow F'=2$ transition is applied to keep the atoms in
the cycling transition for the absorption imaging. The purity of
the samples with respect to Zeeman substates is monitored by
application of a magnetic field gradient during ballistic
expansion of the clouds.

\section{Cross species thermalization}\label{sec:cross}

To measure the interspecies thermalization speed, we perform
experiments at three different temperatures. Each experiment
consists of two stages. First, the mixture is slowly (20$\,$s) cooled to a common
temperature $T_1$. Then, a stage of rapid cooling (50$\,$ms) follows, which leaves
the Rb at a temperature $T_2$, thus producing a temperature difference between the two
atomic ensembles. The thermalization of the Li in the colder Rb
bath is used for determining the scattering length using a straight forward collision model
\cite{Delannoy:01}.  The temperature ranges in the three
 measurements are $T_1/\mu$K$\rightarrow T_2/\mu$K:
$10\rightarrow6$, $6\rightarrow4$ and $4.5\rightarrow2$. On a
timescale of 300$\,$ms thermalization is completed, which is
considerably faster than for $^6$Li, where thermalization times on the order of 3$\,$s were
observed under similar conditions \cite{Silber:05}.

It follows a short description of the thermalization model which
we use to derive the scattering length. The total collision rate
depends on the overlap integral of the two atomic clouds which,
for purely thermal samples in harmonic traps, can be exactly
solved \cite{Mosk:01}
\begin{equation} \Gamma_{coll}=\frac{\sigma_{7,87}}{(2\pi)^{3/2}}\overline{v}N_7N_{87}\left[\frac{m_7\tilde{\omega}_7^2}{k_B(T_{7}+T_{87})}\right]^{3/2}\end{equation}
with the collision cross section $\sigma_{7,87}=4\pi a_{7,87}^2$,
the interspecies pure triplet scattering length $a_{7,87}$, the mean
trap frequency $\tilde{\omega}_7=(\omega_{7,r}^2\omega_{7,z})^{1/3}$ and the mean
relative thermal velocity $\overline{v}=\frac{8k_B}{\pi}\sqrt{\frac{T_7}{m_7}+\frac{T_{87}}{m_{87}}}$.
For atoms with equal masses, an average of 2.7 collisions is
needed for thermalization \cite{Wu:96}, however, in the case of unequal
masses efficiency is reduced by a factor $\xi=\frac{4m_7m_{87}}{(m_7+m_{87})^2}$ \cite{Mosk:01}.
In our case, Li needs
$\sim$10 collisions to thermalize with the Rb. Here we assume the
Rb temperature not to change during the thermalization process.
We ensure this experimentally by application of a constant
frequency microwave knife during the thermalization stage of the
experiment which keeps the Rb temperature (but not the atom number)
fixed. The temperature difference $\Delta T=T_7-T_{87}$ follows
the differential equation
\begin{equation}\frac{d(\Delta T)}{dt}=-\gamma\cdot\Delta T=-\frac{\Gamma_7}{2.7/\xi}\cdot\Delta T\end{equation}
with the collision rate per Li atom $\Gamma_7=\frac{\Gamma_{coll}}{N_7}$.

   \begin{figure}[ht]
        \centerline{\scalebox{0.4}{\includegraphics{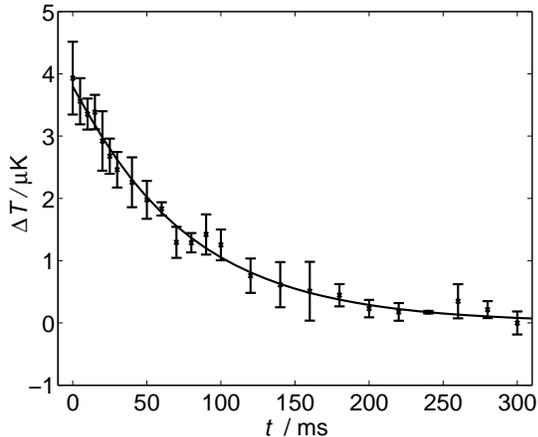}}}
        \caption{Measured (data points) and simulated evolution (solid curve) of the thermalization
        process in the case of $T_1=10\,\mu$K. After about 300$\,$ms the thermalization process is
        complete. Each datapoint is an average of three individual measurements, the errorbars
        indicate the statistical uncertainty.}
        \label{fig_thermalisierung}
    \end{figure}

Application of this model to the measured data results in values
for $|a_{7,87}|$ summarized in table \ref{tab_streulaenge}.

\begin{table}[ht]
\begin{tabular}{c|c|c|c}
$T_1/\mu$K&$T_2/\mu$K&$|a_{7,87}|/a_{\rm B}$&$\chi^2$\\\hline
10 & 6 & 56 & 0.500\\
6 &  4 & 65 & 0.238\\
4.5 & 2 & 56 & 0.924\\
\end{tabular}
\caption{Resulting absolute scattering lengths for three different
thermalization experiments. The quality of the fit $\chi^2$ is
also shown. The variations in the derived scattering length shows
the limitation as to the accuracy in determining $|a_{7,87}|$ with this
method. The mean value is $\overline{a}=59\,a_{\rm B}$ with a standard deviation $\epsilon_{\overline{a}}=5.2\,a_{\rm B}$.
Further errors are discussed in the main text and Fig.\ref{fig_errors}.}\label{tab_streulaenge}
\end{table}

A correction has to be applied to the ballistic expansion time
for the Li component. This is because the self-inductance of the trap coils
results in a coil current shutdown timescale on the order of
200$\,\mu$s. The cloud of Li atoms already starts to expand
during this turn off process whereby the effective time-of-flight is larger
than the nominal value. Hence, the temperatures derived from the
Gaussian widths of the Li clouds using the nominal time-of-flight value are overestimated.
Therefore, an additional experimentally determined correction to
the time-of-flight of 130$\,\mu$s is added to the nominal time of
flight of 1000$\,\mu$s. This changes the measured temperatures by up to 25\,\%
for low temperatures as compared to uncorrected values. For very slow
evaporation ramps, where Li and Rb are in constant equilibrium,
this correction yields the same values for both temperatures. The delayed turn off
effect is negligible for the heavier Rb atoms which we let expand
for a much longer time (20 ms).

\begin{figure}[ht]
        \centerline{
          \scalebox{0.45}{\includegraphics{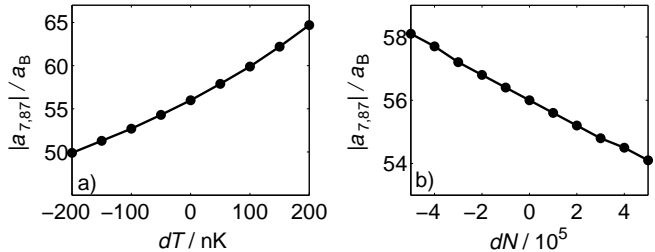}}        }
        \caption{Uncertainty in the temperature difference (a) and Rb atom numbers (b) introduce
         uncertainty in the derived scattering length. Data for $T_1=10\,\mu$K are shown. The values are determined by varying
         the measured $\Delta T$ and $N_{87}$ by small amounts $dT$ and $dN$ and performing the fit to the altered data as in Fig.\ref{fig_thermalisierung}.
         The influence of the uncertainty $dN$ is small compared to that of the uncertainty $dT$. }
        \label{fig_errors}
    \end{figure}

The error introduced due to inexact values for the temperature
difference can be estimated by applying a small temperature offset
to the measured temperature evolution. The effect on the derived
scattering length for $T_1=10\,\mu$K is shown in
Fig.$\,$\ref{fig_errors}a. Mismatched temperature estimates of
only 200$\,$nK in either direction result in variations in $|a_{7,87}|$ of up to
9$\,a_{\rm B}$ for individual $T_1$-measurements. This strongly limits
the accuracy to which we can reliably determine $|a_{7,87}|$.
Another uncertainty derives from the Rb atom numbers, but this
contribution ($\delta |a_{7,87}|/dN\approx0.5\,a_{\rm B}/10^5$) is small
compared to the error introduced due to inexact knowledge of the
Li temperature (Fig.\ref{fig_errors}b). We estimate
$\epsilon_{dT}\approx18\,a_{\rm B}$ and $\epsilon_{dN}\approx1\,a_{\rm B}$.
The total error in $|a_{7,87}|$ is then
$\epsilon_{7,87}=\sqrt{\epsilon_{\overline{a}}^2+\epsilon_{dT}^2+\epsilon_{dN}^2}\approx19\,a_{\rm B}$,
where $\epsilon_{\overline{a}}=5.2\,a_{\rm B}$ is the standard deviation
of the best fit values averaged over the three thermalization
measurements.  Hence, the result of the thermalization experiments
is $|a_{7,87}|=(59\pm19)\,a_{\rm B}$.

 This result does not agree well with calculations based on ab
 initio scattering potentials \cite{Ouerdane:04}, which predict
$a_{7,87}=-0.25\,a_{\rm B}$. With such a value we would not have seen any interspecies
thermalization on the timescale of our experimental cycle.
The deviation is likely to be due to the very
imprecise theoretical knowledge of the scattering potentials. Probably the
good agreement for the measured $^6$Li-$^{87}$Rb scattering length \cite{Silber:05}
with the theoretical value presented in \cite{Ouerdane:04} is coincidental.

\section{Evidence for collapse of the Li component}\label{sec:bec}

\begin{figure}[ht]
        \centerline{\scalebox{0.37}{\includegraphics{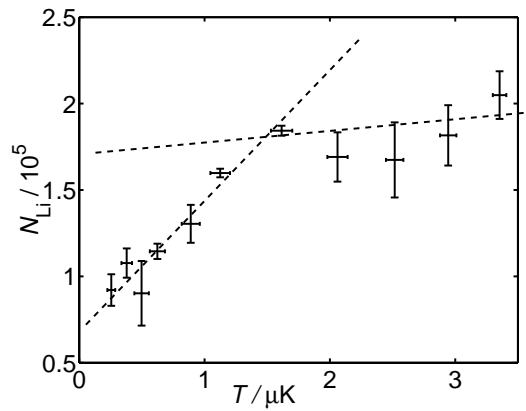}}}
        \caption{The Li atom number drops drastically for temperatures
        lower than $1.5\,\mu$K. The resulting critical atom number
        $N_{c,7}\approx6\cdot10^5$ is larger by a factor of four than
        the measured value. Possible calibration uncertainties are discussed in the text.
        The dashed lines are to guide the eye.}
        \label{fig_collapse}
    \end{figure}

Upon cooling the Rb to even lower temperatures than for the above
thermalization measurements, we reach a regime where we approach
$T_{c}$ for Bose-Einstein condensation of Li. Unlike $^{87}$Rb,
due to their attractive scattering length of
$a_7=-27.3\,a_{\rm B}$, pure condensates of $^7$Li are unstable against
collapse. However, the kinetic pressure in a trapped condensate
allows for a certain maximum number of condensed atoms
\cite{Ruprecht:95,Bergeman:96, Sackett:97}. An estimate of the
maximum atom number in a stable Li condensate for our trapping
geometry gives a value of
$N_{max}\approx\frac{a_{ho,7}}{\left|a_7\right|}\approx700$. Here,
$a_{ho,7}$ is the length scale of the Li harmonic oscillator
ground state. The resolution in detecting atom numbers with our
absorption imaging is insufficient to resolve such small atomic
samples. However, observation of a sudden drop in the Li atom number for
temperatures lower than $1.5\,\mu$K (Fig.\ref{fig_collapse})
provides a means of revealing criticality. Upon reaching the
critical temperature for the Bose-Einstein transition, Li atoms
begin to accumulate in the ground state of the trap, forming a
small condensate. The more atoms condense, the stronger the
condensate is contracted to a size smaller than the harmonic
oscillator groundstate \cite{Ruprecht:95}. Eventually, the
condensate becomes mechanically unstable and collapses upon
itself, thereby strongly increasing the density and thus the rate
of three body recombinations. Depending on the conditions,
collective decay may also be important or even predominant
\cite{Sackett:98,Stoof:97}. All condensed atoms are removed from
the trap because the energy released in these inelastic processes
is much larger than the trap depth. A new condensate is formed on
the timescale of thermalization. With each collapse, this process is
repeated, as long as the thermal cloud of Li atoms is overcritical
with respect to Bose-Einstein condensation. This therefore poses a
permanent loss channel for the much larger number of thermal Li
atoms. Depending on the thermalization rate, the overall decay time can
be on the order of a hundred milliseconds \cite{Adhikari:01b} or
many tens of seconds \cite{Sackett:97}. In our case, we have high
trap frequencies and large thermalization rates. We therefore
expect this process to be rapid.

While the time-of-flight correction for the Li atoms works
well for the temperature range used in the thermalization
experiments, in the low temperature regime we still see a
discrepancy between $T_{87}$ and $T_7$. We observe $T_{87}<T_7$
even though we explicitly took care to ensure thermal equilibrium.
We hence assume $T_7=T_{87}$ for this experiment and plot
the Li data against $T_{87}$ (Fig.\ref{fig_collapse}). With the
approximate value of the measured $T_{c,7}\approx1.5\,\mu$K we
derive a critical atom number of $N_{c,7}=1.2\cdot\left(\frac{k_B
T_{c,7}}{\hbar\tilde{\omega}_7}\right)^3\approx 6\cdot10^5$, which
is a factor of four larger than the measured atom number.

If we assume that the losses are really due to a collapse at $T=1.5\,\mu$K, the
small number of measured atoms needs an explanation. This number can be experimentally
underestimated if the Li atoms do not remain in the
cycling transition used for imaging for the whole duration of the
optical pulse. The cloud therefore becomes transparent
 for the imaging beam yielding less optical density and lower atom numbers.
 A complete description of the
photon absorption process in Li imaging addressing this problem in detail
is under preparation \cite{Deh:07}. Other groups working
with $^6$Li, which, like $^7$Li, suffers from an unresolved hyperfine structure,
image at high magnetic offset fields close to the interesting Feshbach resonance at 822$\,$G
\cite{Zwierlein:04}, where the Paschen-Back effect detunes the overlapping resonances.
Full cycling transitions are possible in this case.

As an alternative to Bose-Einstein criticality, increased losses
could also arise because of strongly energy-dependent three body
loss coefficients that derive from a Feshbach resonance at low
magnetic fields similar to the case of Cs \cite{Weber:03}. This
hypothesis can be safely discarded as a mixture of stretched
states can in principle not experience such interaction
resonances. Yet there may still be a very small fraction of atoms
in the undesired spin states such as $\left|2,1\right>$ below our
detection limit causing residual losses. Even though this cannot
be ruled out, it is more likely that incomplete repumping during
imaging causes incorrect atom number determinations and that
Bose-Einstein criticality is indeed reached \cite{lithium6}.

We have experimentally checked for inelastic losses in the mixture
in the temperature range between $15\,\mu$K and $3\,\mu$K.
We find that the lifetime of Li
in the mixture is reduced for colder temperatures, but still
remains on a timescale of seconds. Going back to the data presented in
Fig.~\ref{fig_collapse}, there, cooling of the mixture from
$4\,\mu$K to any lower value is done in only 500~ms leading to the
observed drastic losses. We therefore conclude that inelastic
collisions at thermal densities cannot account for the sudden
change in behavior of the data below $1.5\,\mu$K unless another
loss channel (condensate collapse) opens. We attribute this loss
channel to be due to the creation and subsequent annihilation of
Li condensates rapidly depleting the observable number of residual
thermal atoms on a short timescale.

\section{Mean-field stabilization}\label{sec:mean}
However, the situation may change dramatically in the presence of
a second \textit{degenerate} species. The mean-field energy can be
either increased or decreased depending on the interaction with
the second species. If the second species itself forms a stable
condensate, as is the case for $^{87}$Rb, it depends on the sign
of the interspecies scattering length, whether the collapse of the
first species is accelerated or slowed down \cite{Adhikari:01}.
For our system this means that for $a_{7,87}<0$ even very small Li
condensates are unstable, as for $a_{7,87}>0$ larger meta stable Li
condensates can be supported.

To demonstrate this we solve the following coupled
Gross-Pitaevskii equation for the spherically symmetric case,
\begin{equation}
    \left[-\frac{\hbar}{2m_k}\frac{1}{r}\frac{d^2}{dr^2}r+\frac{m_k}{2}\tilde{\omega}_k^2r^2+g_kN_k+g_{7,87}N_{l\ne
    k}\right]\psi_k=\mu\psi_k~,
\end{equation}
for $k,l=7\text{ or }87$. Here $g_k=4\pi\hbar^2a_k/m_k$,
$g_{7,87}=2\pi\hbar^2a_{7,87}\frac{m_7+m_{87}}{m_7m_{87}}$ and
$N_{k,l}$ are the atom numbers of both species.

To find the ground state of the system, we use the method of
steepest gradient \cite{Dalfovo:96}. It has already been
successfully employed to study mixed systems of degenerate bosons
and fermions \cite{Roth:02, Modugno:03}. Based on evaluating the
energy functionals $H_k[\psi_k,\psi_l]$, their minimal
values are found by propagating the wave functions along the direction of
the gradient $\frac{\delta H_k}{\delta\psi_k}$ using small time
steps $d\tau$:
\begin{equation}
    \psi_k'\quad\longrightarrow\quad \psi_k+d\tau\,\frac{\delta H_k}{\delta\psi_k}.
\end{equation}
Formally, this corresponds to replacing the real time dependence
in the Gross-Pitaevskii equation by an imaginary time dependence
$\tau=it$. Under the constraint that the normalization of
$\psi_k'$ does not change, which one ensures by explicitly
normalizing $\psi_k'$ after each propagation step, this method is
self-consistent. The final result does not depend on the initial
trial functions, but a sensible choice may significantly reduce
the integration time needed for convergence. We use the ground
state wave functions of the respective harmonic trap potentials as
$\psi_k(\tau=0)$, which are normalized Gaussian functions.

For simplicity we restrict the
calculation to the spherically symmetric case
$V_k(\vec{r})=\frac{1}{2}m_k\tilde{\omega}_k^2 r^2$, where we use
the mean trap frequencies
$\tilde{\omega}_k=(\omega_{k,r}^2\omega_{k,z})^{1/3}$.

The space coordinate was discretized with step size
$dr=0.1\,\mu$m up to $r=40\,\mu$m which corresponds to
$41.9\,a_{ho,87}$ and $22.3\,a_{ho,7}$. Time was discretized using
$d\tau=5\cdot10^{-7}\,$s and the propagation was performed for 6000
time steps.

\begin{figure}[ht]
        \centerline{\scalebox{0.35}{\includegraphics{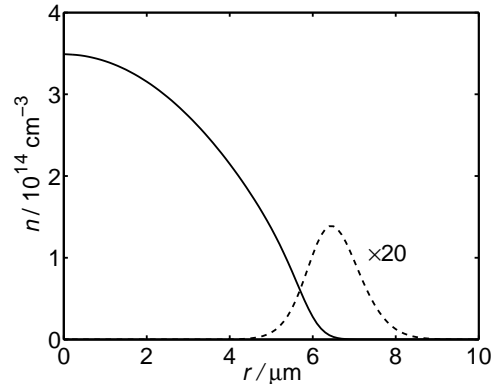}}}
        \caption{Condensate densities of $1.5\cdot10^5$ Rb $\left|2,2\right>$
        (solid line) and 6000 mean-field stabilized Li $\left|2,2\right>$
        atoms (dashed line) with trap parameters from our
        experiment. The interspecies mean-field forces the Li condensate into a shell around
        a large Rb nucleus. The resulting density reduction allows for a much
        larger condensate to survive than in the absence of Rb.
        }\label{fig_coupled_becs}
    \end{figure}

For $a_{7,87}$=0, we get the critical atom number for stable Li
condensates by increasing $N_7$ to the point where we get no
numerically stable solution to the above equation. We find a value
close to the estimated 700 atoms for our trap conditions. Setting
the mutual interaction to repulsive interaction using the measured
absolute value to $a_{7,87}=+59\,a_{\rm B}$ results in a demixing of the
two condensates and a substantial increase in the maximum allowed
number of condensed Li atoms (Fig.\ref{fig_coupled_becs}).

The quoted method only finds the ground state of a given system.
For situations where collapse occurs, it is not suitable to
accurately describe the ongoing complex dynamics. Powerful
numerical schemes based on the Crank-Nicholson scheme
\cite{Holland:95,Dalfovo:00,Adhikari:01} have been developed for studying these
dynamics, but these go beyond the scope of this publication. Here
we restrict ourselves to the static stabilization of Li
condensates in the presence of a Rb condensate. Our results may
experience modifications when considering the cylindrically
symmetric trap employed in our experiment \cite{Esry:97}. Also,
different gravitational sags for the two species may affect the
overall stabilization. This is suggested by experimental results
presented by D.~S. Hall et al. \cite{Hall:98}. They showed that
the mutual interaction between condensates in the two hyperfine
states of Rb caused a spherically symmetric phase-separated state
in the case of vanishing relative gravitational sag. Introduction
of a small gravitational sag of $0.4\,\mu$m, which was smaller
than the Thomas-Fermi radii of the condensates involved, caused
strongly damped dynamics and led to a vertically phase-separated
state with only limited spatial overlap between the two
condensates. Calculations incorporating the effect of gravity will
have to be performed to clarify whether or not mean-field
stabilization then is still significant. In order to ensure
symmetry as much as possible, tightly confining traps with
strongly reduced relative gravitational sag should be employed to
study mean-field stabilization.

\bigskip

Based on the hypothesis that $a_{7,87}$ is positive, we
performed a series of experiments where both Li and Rb are slowly
cooled to about $3\,\mu$K which is well above the experimentally
determined value of the critical temperature for Li condensation.
The respective critical temperatures behave as
$T_{c,7}=\sqrt{\frac{87}{7}}\left(\frac{N_{87}}{N_{7}}\right)^{1/3}\,T_{c,87}$.
The large mass ratio and the therefore strongly different trap
frequencies lead to a higher critical temperature for Li given our
experimental conditions, i.e. $T_{c,7}\approx1.5\,\mu$K while
$T_{c,87}\approx0.7\,\mu$K. In order to generate a Rb condensate
without losing too many Li atoms during the evaporation to
$T_{c,87}$, we very quickly (within 200$\,$ms) cool the Rb
component into degeneracy. The evaporating microwave knife remains activated
at the end of the evaporation ramp resulting in a slow temperature
decrease from initially $800\,$nK down to about $200\,$nK during
the 900~ms storage time. During the
storage stage Rb condensates with atom numbers ranging from
some $10^4$ up to $1.5\cdot10^5$ atoms are formed. The
atom number of the Li component drops sharply to less than $10^4$
at the end of the 900$\,$ms and no bimodal density distribution,
the typical sign for Bose-Einstein condensation, was observed at
any time for the Li component. Extraction of the
Li temperature from the time-of-flight images renders values
around $1\,\mu$K throughout the duration of the experiment while
the collisional thermalization model suggests that the thermalization is very
fast so that $\Delta T<100\,$nK for $t>200~$ms. Specifically, the
amount of Rb atoms during the experiment is sufficient to ensure
thermalization \textit{even} with an underestimation of the Li
atom number by a factor of four.
 We therefore believe that the Li temperatures derived
from Li absorption imaging at low temperatures are not reflecting the
actual temperature of the species. Accurate values for very low Li temperatures seem to be not
achievable under current conditions. A conclusion as to the sign
of the scattering length is at this stage not possible. Two possible scenarios remain:
\begin{enumerate}\item The sign is positive, but the relative gravitational sag
reduces the overlap so that stabilization is suppressed.
The Li condensates stay below the detection limit. \item The sign
is negative and the Li condensates are made even more unstable
rendering them unresolvable for our system.
\end{enumerate}
To reduce the effects of gravity on the overlap of the atomic clouds, one should
perform these experiments again in much tighter confinement. The different
trap geometry, cigar shaped as opposed to spherically symmetric, has to be considered in the
theoretical description as well, the get a better quantitative picture of mean-field stabilization
under our experimental conditions.

\section{Conclusion and outlook}

 In conclusion, we cooled a mixed gas of two
atomic species, $^7$Li and $^{87}$Rb down to ultracold
temperatures. We determined the absolute value of the pure triplet
interspecies scattering $|a_{7,87}|$ through measurements of the
thermalization speed to be $59\pm19$ Bohr radii. The cause of
slow atom losses experienced by the Li during the cooling process
remains unclear. As Li and Rb atoms in other states than
$\left|2,2\right>$ are kept below our detection limit direct
inelastic loss processes can be ruled out. Sudden severe Li losses
upon cooling below a critical temperature are interpreted as the
onset of Bose-Einstein condensation for the Li component. Even
though numerical mean-field calculations suggest increased
stabilization of the Li condensate in case of repulsive
interspecies interaction, Li condensates in the presence of a Rb
condensate were not detected even in the case of large Rb
condensates. Such an observation directly determines the sign of
the interspecies interaction to be positive whereas from the
current observational viewpoint no decisive conclusion as to the
sign of $a_{7,87}$ can be made. The effects of the different trap
geometry and of gravitational sag need to be taken into account
for more accurate quantitative predictions. More tightly confining
traps that reduce the effect of gravity may then prove to be more
suitable to study mean-field stabilization. A detailed study of the
imaging process of Li at low magnetic fields is under preparation addressing the
atom number inconsistencies \cite{Deh:07}.

\bigskip

This work has been supported by the Deutsche
Forschungsgemeinschaft (DFG).


\begin{thebibliography}{10}
\bibitem{interacting_mots} As the two magneto-optical traps strongly interact inelastically in the short compressed stage, the Li atom number
                           can be a factor of 4 higher when we do not load Rb.
\bibitem{lithium6}We experienced similar Li atom number inconsistencies in our study
      of the fermionic species $^6$Li when considering Fermi degeneracy. $^6$Li also has an unresolved hyperfine structure of the excited $2^3P_1$ state.
\end{thebibliography}
\end{document}